\documentclass[showpacs,10pt]{article}
\usepackage{eurosym}
\usepackage{graphicx,color}
\usepackage{amsmath}
\usepackage{amssymb}
\usepackage{amsmath,amsfonts,latexsym,graphicx,amssymb}
\usepackage{amsfonts}
\usepackage{amssymb}
\usepackage{mathrsfs}
\usepackage{amsthm}
\usepackage{graphicx}
\usepackage{amsmath}

\newcommand{{\Cd}}{{\mathbb{C}^d}}
\newcommand{{\Rn}}{{\mathbb{R}^n}}

\def\<{\langle}
\def\>{\rangle}
\newtheorem{thm}{Theorem}

\newtheorem{ex}{Example}

\newtheorem{pro}{Proposition}

\begin{document}

\date{}
\title{\textbf{Stochastic evolution of finite level systems: \\
classical vs. quantum}}
\author{D. Chru\'sci\'nski$^1$\thanks{%
email: darch@fizyka.umk.pl}, V.I. Man'ko$^2$, G. Marmo$^{3,4}$, and F.
Ventriglia$^{3,4}$ \\
\\
$^1$Institute of Physics, Faculty of Physics, Astronomy and Informatics, \\
Nicolaus Copernicus University, Grudzi{a}dzka 5, 87--100 Toru\'n, Poland \\
\\
$^2$P. N. Lebedev Physical Institute, Russian Academy of Sciences, 119991
Moscow, Russia \\
\\
$^3$Dipartimento di Fisica and MECENAS, \\
Universit\`a di Napoli Federico II, I-80126
Napoli, Italy \\
\\
$^4$INFN, Sezione di Napoli, I-80126 Napoli, Italy}
\maketitle

\begin{abstract}
Quantum dynamics of the density operator in the framework of a single
probability vector is analyzed. In this framework quantum states define a
proper convex \emph{quantum} subset in an appropriate simplex. It is showed
that the corresponding dynamical map preserving a quantum subset needs not
be stochastic contrary to the classical evolution which preserves the entire
simplex. Therefore, violation of stochasticity witnesses quantumness of
evolution.

\end{abstract}

\section{Introduction}

The basic idea of quantum tomography is to encode the information of a
quantum state in a family of probability distributions. This idea goes back
to Pauli \cite{Pauli} who considered the possibility to associate quantum
states with marginal probability distributions in the phase space, both for
positions and momenta, as in classical statistical mechanics. In spite that
these marginal distributions do not determine the quantum state wave
function \cite{Reichenbach}, that idea was realized by considering an
infinite family of marginal distributions (tomograms).

The tomographic representation is based on the Radon transform of the Wigner
function which relates the measurable optical tomographic probability to
reconstruct the Wigner function of the quantum state (see \cite{Beppe} for a
recent review). Similar tomographic approach may be constructed for finite
level quantum systems (spin systems) corresponding to finite-dimensional
Hilbert space $\mathcal{H}$ \cite{D1,D2,D3}. In this case a spin (or qudit)
density operator $\rho $ can be described \cite{Tom1} by probability
distributions of random spin projection (spin tomogram) depending on the
direction of the quantization axis. Such family of probability distribution
encodes the complete information about qudit state and hence a density
operator may be uniquely recovered out of a spin tomogram.

Quantum tomography enables one to encode the information about $\rho $ into
the finite family of probability distributions (see the next section) and
eventually to represent $\rho $ as a single probability vector living in an
appropriate simplex. It is clear that only a convex \emph{quantum subset}
within this simplex correspond to legitimate density operators. Using this
stochastic representation $\rho \rightarrow \mathbf{P}$ one may translate a
quantum evolution of $\rho (t)$ into a stochastic evolution of $\mathbf{P}(t)
$. The problem of stochastic evolution and dynamical maps of density
matrices was studied and reviewed recently in \cite{Chru1,Chru2,Chru3}. In this way both classical
and quantum evolution of finite level systems may be described within the
same framework. Now comes a natural question: suppose that one is given the
evolution of a stochastic vector $\mathbf{P}(t)=\mathbb{T}(t)\mathbf{P}$
represented by a semigroup of maps $\mathbb{T}(t)$ satisfying $\mathbb{T}%
(t+u)=\mathbb{T}(t)\mathbb{T}(u)$ for all $t,u\geq 0$. Is it possible to
discriminate wether $\mathbb{T}(t)$ describes purely classical evolution of $%
\mathbf{P}$ or a quantum evolution of $\rho $ encoded in $\mathbf{P}$? It is
clear that classical dynamics is governed by stochastic matrices $\mathbb{T}%
(t)$. Interestingly, it needs be no longer true for quantum dynamics. Some
preliminary aspects of this phenomenon were mentioned in \cite{Fil,Tras} Note
that in the quantum case one requires that $\mathbb{T}(t)$ maps probability
vectors $\mathbf{P}$ within the \emph{quantum subset} of the simplex. In
particular it turns out that if the dynamics of $\rho (t)$ is unitary then $%
\mathbb{T}(t)$ is never a stochastic family of maps. This way violation of
stochasticity witnesses quantumness of evolution.

The paper is organized as follows. In Section \ref{TOM} we review the
tomographic approach to quantum mechanics. In Section \ref{P} we show how to
encode the information about a density operator $\rho$ into a single
probability vector $\mathbf{P}$ living in an appropriate simplex. Vectors
corresponding to legitimate quantum states form a convex \emph{quantum subset%
} within a simplex. It is shown how the evolution of $\rho(t)$ induces the
evolution of $\mathbf{P}(t)$. Finally, Section \ref{EX} provides several
examples of qubit dynamics in the framework of 6-dimensional probability
vector $\mathbf{P}$. Final conclusions are collected in the last Section.


\section{Tomography of spin states}

\label{TOM}

In this section we review the approach \cite{Beppe} to describe the quantum
states of a spin (qudit) by means of the probability distributions called
quantum tomograms. The tomogram $W(m|U)$ corresponding to a qudit state with
the density operator $\rho$ is given by the diagonal matrix elements of the
unitarily rotated density operator in the standard computational basis $|m\>$%
, where $-j \leq m \leq j$ and $J_z |m\> = m|m\>$, with $J_z$ being a spin
projection operator on $z$-axis. Thus one has
\begin{equation}
W(m,U) = \<m|U \rho \,U^\dagger|m\> \ ,
\end{equation}
where $U$ is a $(2j+1)\times (2j+1)$ unitary matrix. In the case when the
matrix $U$ is a matrix of irreducible representation of the group $SU(2)$
the tomogram $W(m|\mathbf{n})$, called a spin tomogram, introduced in \cite{D1,D2}
depends on the variable $\mathbf{n}=(\sin\theta\cos\varphi,\sin\theta\sin%
\varphi,\cos\theta)$. Spherical angles $(\theta,\varphi)$ determine the
point on the Poincar\'e sphere. Both tomograms $W(m|U)$ and $W(m|\mathbf{n})$
define probability distributions, that is, $W(m,U)\geq 0$, $W(m,\mathbf{n}%
)\geq 0$, and the normalization condition
\begin{equation}
\sum_{m=-j}^j W(m,U) = 1\ , \ \ \ \ \sum_{m=-j}^j W(m,\mathbf{n}) = 1\ .
\end{equation}
These properties mean that the tomograms can be
interpreted as conditional probability distributions $W(m,U) \equiv W(m|U)$
and $W(m,\mathbf{n})\equiv W(m|\mathbf{n})$ which can be associated with a
joint probability distributions $\mathcal{P}(m,U)$ or $\mathcal{P}(m,\mathbf{%
n})$. The construction of the probability distribution $\mathcal{P}(m,%
\mathbf{n})$ in the form $\mathcal{P}(m,\mathbf{n}) = W(m,\mathbf{n}) \Pi(%
\mathbf{n})$, where $\Pi(\mathbf{n})$ is a probability distribution on the
sphere $S^2,$ was discussed for particular cases in \cite{albini,BelFil} and the probability vector with
components $W(m,U_j)$ ($j=1,\ldots,N+1)$ for qudit states was considered in
\cite{aip}. The important property of this probability vector is that
for the corresponding density operator $\rho$ describing the quantum states
the vector components occupy only a subset of the corresponding simplex. To
see this let us consider the conditional probability
\begin{equation}
p(m|\alpha) = \<m |U_\alpha \rho \, U_\alpha^\dagger| m \>\ ,
\end{equation}
for $m=1,\ldots,N$ and $\alpha=1,\ldots,N+1$ (with $N=2j+1$). One assumes
that the set of unitaries $\{U_1,\ldots,U_{N+1}\}$ defines a quorum, that
is, one may reconstruct $\rho$ out of $p(m|\alpha)$. Note that $p(m|\alpha)$
encodes $(N+1)\times (N-1)=N^2-1$ real parameters, that is, exactly the same
number as the density operator $\rho$ in $N$-dimensional complex Hilbert
space. Fixing the $(N+1)$-dimensional probability distribution $\pi_\alpha$,
that is, we take into account the measurement of $U_\alpha$ with a weight $%
\pi_\alpha >0$, one defines
\begin{equation}
p^{(\alpha)}_k := p(k|\alpha) \pi_\alpha\ ,
\end{equation}
and finally $N(N+1)$-dimensional probability vector
\begin{equation}
\mathbf{P} = (p^{(1)}_1, \ldots, p^{(1)}_N, \ldots,
p^{(N+1)}_1,\ldots,p^{(N+1)}_N)^{\mathrm{T}}\ .
\end{equation}
It is clear that only very special vectors $\mathbf{P}$ correspond to
legitimate quantum states on $N$-level quantum system. Any such vector has
to respect $(N+1)$ linear constraints
\begin{equation}
p^{(\alpha)}_1 + \ldots + p^{(\alpha)}_N = \pi_\alpha\ ,
\end{equation}
for each $\alpha=1,2,\ldots,N+1$. It is clear that a proper set of $U_\alpha$
enables one to reconstruct $\rho$ out of $\mathbf{p}^{(\alpha)}$. Note, that
positivity of $\rho$, that is, $\rho \geq 0$, provides highly nontrivial
constraint for a set probabilities $\mathbf{p}^{(\alpha)}$ and hence for a
set of admissible vectors $\mathbf{P}$. Hence, probability vectors $\mathbf{P%
}$ corresponding to legitimate quantum states define only a proper $(N^2-1)$%
--dimensional convex subset $Q_N$ of $(N^2+N-1)$--dimensional simplex $%
\Sigma_{N(N+1)}$.

To summarize: using a tomographic approach with a fixed quorum $U_\alpha$
any quantum state $\rho$ of $N$-level quantum system may be attributed a
probability vector $\mathbf{P}$ in $N(N+1)$-simplex. Vectors compatible with
legitimate quantum states define a proper convex subspace $Q_N$.

As an illustration consider a qubit state $\rho$. Using the Bloch
representation one has
\begin{equation}
\rho = \frac 12 \left( \mathbb{I}_2 + x\sigma_x + y\sigma_y + z\sigma_z
\right) \ .
\end{equation}
The requirement $\rho \geq 0$ provides the following constraint for the
Bloch vector $\mathbf{r} =(x,y,z)$:
\begin{equation}  \label{Bloch}
|\mathbf{r}|^2 = x^2 + y^2 + z^2 \leq 1\ .
\end{equation}
The corresponding conditional probability $p(m|\alpha)$ reads:
\begin{eqnarray*}
p(1|x) &=& \frac{1+x}{2} \ , \ \ \ p(2|x) = \frac{1-x}{2} \ , \\
p(1|y) &=& \frac{1+y}{2} \ , \ \ \ p(2|y) = \frac{1-y}{2} \ , \\
p(1|z) &=& \frac{1+z}{2} \ , \ \ \ p(2|z) = \frac{1-z}{2} \ ,
\end{eqnarray*}
where as a quorum one takes $\{ U_x=\sigma_x,U_y=\sigma_y,U_z=\sigma_z\}$.
Hence fixing a probability distribution $(\pi_x,\pi_y,\pi_z)$ one defines
\begin{equation}
\mathbf{P} = \left( p^{(x)}_1, p^{(x)}_2, p^{(y)}_1, p^{(y)}_2, p^{(z)}_1,
p^{(z)}_2 \right)^{\mathrm{t}}\ .
\end{equation}
One easily finds the following relations:
\begin{eqnarray*}
p^{(x)}_1 &=& \frac{\pi_x}{2} \big( 1 + [\rho_{12} + \rho_{21}] \big) \ , \
\ \ p^{(y)}_1 = \frac{\pi_y}{2}\big( 1 - i[\rho_{12} - \rho_{21}] \big)\ , \
\ \ p^{(z)}_1 = \pi_z \rho_{11}\ , \\
p^{(x)}_2 &=& \frac{\pi_x}{2}\big( 1 - [\rho_{12} + \rho_{21}] \big)\ , \ \
\ p^{(y)}_2 = \frac{\pi_y}{2}\big( 1 + i[\rho_{12} - \rho_{21}] \big)\ , \ \
\ p^{(z)}_2 = \pi_z \rho_{22} \ .
\end{eqnarray*}
Vectors $p^{(\alpha)}_k$ satisfy 3 linear constraints
\begin{equation}
p^{(x)}_1 + p^{(x)}_2 = \pi_x\ , \ \ \ p^{(y)}_1 + p^{(y)}_2 = \pi_y\ , \ \
\ p^{(z)}_1 + p^{(z)}_2 = \pi_z\ .
\end{equation}
The quadratic constraint (\ref{Bloch}) yields
\begin{equation}  \label{Bloch-ell}
\frac{[ p^{(x)}_1 - a_x ]^2}{a_x^2} + \frac{[ p^{(y)}_1 - a_y]^2}{a_y^2} +
\frac{ [ p^{(z)}_1 - a_z]^2}{a_z^2} \leq 1\ ,
\end{equation}
where
\begin{equation}
a_x = \frac{\pi_x}{2}\ , \ \ \ a_y = \frac{\pi_y}{2}\ , \ \ \ a_z = \frac{%
\pi_z}{2}\ .
\end{equation}
Hence a Bloch ball (\ref{Bloch}) is transformed into an ellipsoid (\ref%
{Bloch-ell}). Note that for uniform distribution $\pi_x=\pi_y=\pi_z = \frac
13$ the above ellipsoid reduced to a ball.

\section{Classical vs. quantum dynamics of probability vectors}

\label{P}

Both classical and quantum states of finite levels systems can be encoded as
probability vectors in the corresponding simplex. Now comes a crucial
question: suppose we are given a probability vector $\mathbf{P}_0$ and its
evolution $\mathbf{P}(t)$ for $t \geq 0$. Can we discriminate between
classical and quantum evolution?

Let us briefly recall standard description of classical and quantum
stochastic dynamics. Consider first an $N$-level classical systems described
by the following master equation
\begin{equation}  \label{MEC}
\frac{d}{dt} \mathbf{p}(t) = M \mathbf{p}(t)\ , \ \ \ \mathbf{p}(0)=\mathbf{p%
}_0\ ,
\end{equation}
where $\mathbf{p}= (p_1,\ldots,p_N)^{\mathrm{t}}$ denotes probability
distribution, and $M$ stands for the time-independent generator. The
solution defines a classical dynamical map $T(t)$ such that $\mathbf{p}(t) =
T(t)\mathbf{p}_0$. It is clear that $T(t)$ is a stochastic $N\times N$
matrix such that $T(0)=\mathbb{I}_N$. Let us recall that a real $n\times n$
matrix $T$ is stochastic iff: i) $T_{ij} \geq 0$, and ii) $\sum_i T_{ij}=1$
for all $j=1,\ldots,n$. The master equation (\ref{MEC}) rewritten in terms
of $T(t)$ reads
\begin{equation}  \label{MEC}
\frac{d}{dt} T(t) = M \, T(t)\ , \ \ \ T(0)= \mathbb{I}_N\ ,
\end{equation}
and hence the solution is given by $T(t) = e^{t M}$. It is well known \cite%
{Kampen} that $e^{t M}$ provides a stochastic matrix for all $t \geq 0$ if
and only if $M$ satisfies the following Kolmogorov conditions : i) $M_{ij}
\geq 0$ for $i \neq j$, and ii) $\sum_i M_{ij} = 0$ for all $j=1,\ldots,N$.
In this case $M$ possesses the following representation
\begin{equation}
M_{ij} = \pi_{ij} - \delta_{ij} \sum_{k=1}^n \pi_{kj} \ ,
\end{equation}
with $\pi_{ij} \geq 0$ for $i \neq j$. Actually, the diagonal elements $%
\pi_{ii}$ cancel out and hence they are completely arbitrary. Putting this
form into (\ref{MEC}) one arrives at the classical Pauli rate equation
\begin{equation}
\frac{d}{dt}\, p_i(t) = \sum_{j=1}^N \left[\, \pi_{ij} p_j(t) - \pi_{ji}
p_i(t)\, \right] \ ,
\end{equation}
where $\pi_{ij}$ play a role of transition rates.

Consider now the dynamics of a quantum $n$-level system. The corresponding
quantum master equation reads
\begin{equation}  \label{MEQ}
\frac{d}{dt} \rho(t) = L \rho(t)\ , \ \ \ \rho(0)=\rho_0\ ,
\end{equation}
where $\rho$ denotes a density operator, and $L$ stands for the
time-independent (quantum) generator. The corresponding quantum evolution
gives rise to the family of dynamical maps $\Lambda(t) : \mathfrak{T}(%
\mathcal{H}) \rightarrow \mathfrak{T}(\mathcal{H})$ such that $\rho(t) =
\Lambda(t)\rho_0$. For any $t\geq 0$ a map $\Lambda(t)$ is completely
positive and trace preserving (i.e. defines a quantum channel). The formal
solution $\Lambda(t) = e^{tL}$ defines legitimate dynamical map if and only
if the corresponding generator has a well known
Gorini-Kossakowski-Sudarshan-Lindblad form \cite{GKS,Lindblad} (see also
\cite{Alicki})
\begin{equation}
L\rho = -i[H,\rho] + \frac 12 \sum_k \Big( [V_k\rho,V_k^\dagger] + [V_k,\rho
V_k^\dagger] \Big)\ ,
\end{equation}
where $H^\dagger = H \in \mathfrak{B}(\mathcal{H})$ and $V_k \in \mathfrak{B}%
(\mathcal{H})$ are arbitrary. Interestingly, if one fixes an orthonormal
basis $\{e_1,\ldots,e_n\}$ in $\mathcal{H}$, then the following matrix
\begin{equation}
M_{ij} = \mathrm{Tr}[ P_i L(P_j) ]\ ,
\end{equation}
where $P_i = |e_i\>\<e_i|$, satisfies Kolmogorov conditions, and
\begin{equation}
T_{ij}(t) = \mathrm{Tr}[ P_i \Lambda(t)(P_j) ]\ ,
\end{equation}
is stochastic for all $t\geq 0$.

Up to now we described classical and quantum evolutions using different
frameworks: stochastic evolution of a probability vector and completely
positive evolution of a density operator. To compare how classical and
quantum system evolve let us encode $\rho$ as a $n(n+1)$--dimensional
probability vector $\mathbf{P} \in Q_n$. It is clear that the quantum master
equation (\ref{MEQ}) rewritten in terms of $\mathbf{P}$ gives rise to the
following linear equation
\begin{equation}  \label{MEQn}
\frac{d}{dt}\, \mathbf{P}(t) = \mathbb{M}\, \mathbf{P}(t) \ ,
\end{equation}
and the corresponding solution has the following form $\mathbf{P}(t) =
\mathbb{T}(t)\mathbf{P}_0$, with $\mathbb{T}(t) = e^{t \mathbb{M}}$.

Now comes the question: suppose one is given a master equation (\ref{MEQn})
for a probability vector $\mathbf{P}\in Q_n \subset \Sigma_{n(n+1)}$ Is it
possible to discriminate whether $\mathbf{P}(t)$ represents classical
dynamics of $n(n+1)$-level system or quantum dynamics of $n$-level system
encoded in $\mathbf{P}$? Any $\mathbb{M}$ satisfying Kolmogorov conditions
gives rise to legitimate classical dynamics. It means that in the classical
case $\mathbb{T}(t) = e^{t \mathbb{M}}$ is always stochastic matrix for $t
\geq 0$. Interestingly, for quantum dynamics it is no longer the case. It
means that if $\mathbb{T}(t)$ represents quantum evolution it needs not be a
family of stochastic maps. Equivalently, the corresponding generator needs
not have Kolmogorov form. Hence, whenever $\mathbb{T}(t)$ violates
stochasticity it proves the quantumness of the corresponding evolution.

\section{Examples: qubit dynamics}

\label{EX}

In this section we provide simple examples of qubit dynamics giving rise to the
evolution $\mathbb{T}(t)$ which needs not be stochastic.

\begin{pro}
If $L\rho = - i[H,\rho]$, then the corresponding $\mathbb{M}$ is never of
Kolmogorov form.
\end{pro}

Indeed, note that in this case the dynamics is invertible, that is, $L$ and $%
-L$ share the same property. Now, if $\mathbb{M}$ were of Kolmogorov form
then $-\mathbb{M}$ is not and hence the symmetry is broken. Note, that $%
M_{ij} = \mathrm{tr}[P_i L(P_j)] \equiv 0$.

\begin{ex}
\emph{Consider a Schr\"odinger evolution of a qubit state governed by the
von Neumann equation
\begin{equation}
\frac{d}{dt}\rho(t) = -i [H,\rho(t)]\ ,
\end{equation}
with $H= \omega \sigma_x$. One finds
\begin{equation}
\mathbb{M} = \omega \left(
\begin{array}{cc|cc|cc}
0 & 0 & 0 & 0 & 0 & 0 \\
0 & 0 & 0 & 0 & 0 & 0 \\ \hline
0 & 0 & 0 & 0 & -2 \,\mu & 2\, \mu \\
0 & 0 & 0 & 0 & 2\, \mu & -2\, \mu \\ \hline
0 & 0 & \mu^{-1} & - \mu^{-1} & 0 & 0 \\
0 & 0 & - \mu^{-1} & \mu^{-1} & 0 & 0%
\end{array}
\right) \ ,
\end{equation}
with $\mu = \pi_y/\pi_z$. It is clear that $\mathbb{M}$ is not a Kolmogorov
generator and hence the evolution $\mathbb{T}(t)=e^{t \mathbb{M}}$ is not
stochastic. We stress that a sign of $\omega$ does not play any role. }
\end{ex}

One might be tempted to expect that purely dissipative quantum dynamics
leads to legitimate stochastic evolution $\mathbb{T}(t)$. To show that it is
not the case let us consider an elementary generator constructed out of a
single operator $V$:
\begin{equation}
L \rho = V \rho V^\dagger - \frac 12 \big( V^\dagger V \rho + \rho V^\dagger
V \big)\ .
\end{equation}
One finds for the diagonal elements
\begin{eqnarray*}
\dot{\rho}_{11} &= & - \gamma \rho_{11} + \gamma \rho_{22} + \kappa\,
\rho_{12} + \overline{\kappa}\, \rho_{21} \ , \\
\dot{\rho}_{22} &=& \ \ \, \gamma \rho_{11} - \gamma \rho_{22} - \kappa\,
\rho_{12} - \overline{\kappa}\, \rho_{21} \ ,
\end{eqnarray*}
where
\begin{equation*}
\gamma = |V_{12}|^2\ , \ \ \ \kappa = V_{11} V^\dagger_{21} - \frac 12
\<2|V^\dagger V|1\> = \frac 12 ( V_{11} V^\dagger_{21} - V_{21}
V^\dagger_{22} ) \ ,
\end{equation*}
and hence
\begin{eqnarray*}
\frac{d}{dt}\, p^{(z)}_1 &= & - \gamma p^{(z)}_1 + \gamma p^{(z)}_2 + X_1
p^{(x)}_1 + X_2 p^{(x)}_2 + Y_1 p^{(y)}_1 + Y_2 p^{(y)}_2 \ , \\
\frac{d}{dt}\, p^{(z)}_2 &=& \ \ \, \gamma p^{(z)}_1 - \gamma p^{(z)}_2 -
X_1 p^{(x)}_1 - X_2 p^{(x)}_2 - Y_1 p^{(y)}_1 - Y_2 p^{(y)}_2 \ ,
\end{eqnarray*}
for real parameters $X_1,X_2,Y_1,Y_2$ (they are easily calculable in terms
of $\kappa$ and $\overline{\kappa}$). It is therefore clear that unless $%
X_1=X_2=Y_1=Y_2=0$ the corresponding dynamics of probability vector $\mathbf{%
P}$ can not be stochastic: indeed if $X_1 >0$ then $-X_1<0$ and hence the
generator does not satisfies Kolmogorov conditions $\mathbb{M}_{ij} \geq 0$
for $i \neq j$. The above analysis gives rise to the following

\begin{thm}
A legitimate quantum evolution $\Lambda(t)$ gives rise to legitimate
stochastic evolution $\mathbb{T}(t)$ if and only if $\mathbb{T}(t)$ is block
diagonal, that is,
\begin{equation}
\mathbb{T}(t) = \bigoplus_\alpha \, {T}^{(\alpha)}(t) \ ,
\end{equation}
and each $\mathbb{T}^{(\alpha)}(t)$ defines stochastic evolution of the
probability distribution $p(m|\alpha) = p^{(\alpha)}_m/\pi_\alpha$.
Equivalently, the corresponding generator
\begin{equation}
\mathbb{M} = \bigoplus_\alpha \, {M}^{(\alpha)} \ ,
\end{equation}
and each ${M}^{(\alpha)}$ satisfies Kolmogorov conditions.
\end{thm}

Interestingly, several well known quantum generators leads to legitimate
Kolmogorov form of $\mathbb{M}$.

\begin{ex}
\emph{Consider a dissipative evolution governed by the following master
equation
\begin{eqnarray}
\frac{d}{dt}\rho(t) = -i \frac{\omega}{2} [\sigma_z,\rho(t)] + L_D \rho(t) \
,
\end{eqnarray}
where the dissipative part $L_D$ is defined by
\begin{equation}
L_D \rho = \frac 12 \Big\{ \gamma_1 \big([\sigma_+\rho,\sigma_-] +
[\sigma_-,\rho\sigma_+]\big) + \gamma_2 \big([\sigma_-\rho,\sigma_+] +
[\sigma_+,\rho\sigma_-] \big) + \gamma_3 \big( \sigma_z \rho\sigma_z - \rho %
\big) \Big\} \ ,
\end{equation}
with $\sigma_+=|2\>\<1|$ and $\sigma_-=|1\>\<2|$ the standard raising and
lowering operators. One easily find the corresponding generator
\begin{equation}
\mathbb{M} = \frac 12 \left(
\begin{array}{cc|cc|cc}
-\Gamma & \Gamma & -\omega\nu & \omega\nu & 0 & 0 \\
\Gamma & -\Gamma & \omega\nu & -\omega\nu & 0 & 0 \\ \hline
\omega/\nu & -\omega/\nu & -\Gamma & \Gamma & 0 & 0 \\
-\omega/\nu & \omega/\nu & \Gamma & -\Gamma & 0 & 0 \\ \hline
0 & 0 & 0 & 0 & -2{\gamma}_1 & 2{\gamma}_2 \\
0 & 0 & 0 & 0 & 2{\gamma}_1 & -2{\gamma}_2%
\end{array}
\right) \ ,
\end{equation}
where $\Gamma = \frac 12 (\gamma_1 + \gamma_2) + \gamma_3$ and $\nu =
\pi_x/\pi_y$. Interestingly, block-diagonal blocks satisfy separately
Kolmogorov conditions. Note that if $\omega=0$, i.e. dynamics is purely
dissipative, then $\mathbb{M}$ is block-diagonal and hence perfectly of the
Kolmogorov form. }
\end{ex}

\begin{ex}
\emph{Consider the following generator
\begin{equation}
L \rho = \sum_{k=1}^3 \gamma_k \big( \sigma_k \rho \sigma_k - \rho \big)\ ,
\end{equation}
with $\gamma_k \geq 0$. Evolution generated by the above generator is called
random unitary since it is defined by the following formula
\begin{equation}
\Lambda(t)\rho = \sum_{\mu=0}^3 \, p_\mu(t)\, \sigma_\mu \rho \, \sigma_\mu\
,
\end{equation}
where $p_\mu(t)$ is a probability vector for $t \geq 0$. One finds
\begin{eqnarray*}
p_0(t) &=&\frac{1}{4}\, [1+ A_{12}(t) + A_{13}(t) + A_{23}(t)] \ , \ \ \
p_1(t) =\frac{1}{4}\, [1- A_{12}(t) - A_{13}(t) + A_{23}(t)] \ , \\
p_2(t) &=&\frac{1}{4}\, [1- A_{12}(t) + A_{13}(t) - A_{23}(t)] \ ,\ \ \
p_3(t) =\frac{1}{4}\, [1+ A_{12}(t) - A_{13}(t) - A_{23}(t)] \ ,
\end{eqnarray*}
where $A_{ij}(t) = e^{-2[(\gamma_i + \gamma_j)t]}$. Using the following
matrix representation
\begin{equation*}
L \rho = \left(
\begin{array}{cc}
(\gamma_1 +\gamma_2)(\rho_{22}-\rho_{11}) & - \rho_{12}( \gamma_1 + \gamma_2
+ 2 \gamma_3) + \rho_{21}( \gamma_1 - \gamma_2) \\
- \rho_{21}( \gamma_1 + \gamma_2 + 2 \gamma_3) + \rho_{12}( \gamma_1 -
\gamma_2) & (\gamma_1 +\gamma_2)(\rho_{11}-\rho_{22})%
\end{array}
\right)\ ,
\end{equation*}
one easily finds
\begin{equation}
\mathbb{M} = M^{(x)} \oplus M^{(y)} \oplus M^{(z)} \ ,
\end{equation}
with
\begin{equation}
M^{(x)} = \left(
\begin{array}{cc}
- \gamma_x & \gamma_x \\
\gamma_x & - \gamma_x%
\end{array}
\right) \ , \ \ \ M^{(y)} = \left(
\begin{array}{cc}
- \gamma_y & \gamma_y \\
\gamma_y & - \gamma_y%
\end{array}
\right) \ , \ \ \ M^{(z)} = \left(
\begin{array}{cc}
- \gamma_z & \gamma_z \\
\gamma_z & - \gamma_z%
\end{array}
\right) \ ,
\end{equation}
and
\begin{equation*}
\gamma_x = \gamma_2 + \gamma_3 \ ,\ \ \ \gamma_y = \gamma_1 + \gamma_3 \ ,\
\ \ \gamma_z = \gamma_1 + \gamma_2 \ .
\end{equation*}
Hence random unitary dynamics of a qubit provides perfectly stochastic
dynamics of the 6-dimensional probability vectors $\mathbf{P}$.
Interestingly, one has three 2-dimensional stochastic evolutions
\begin{equation}
\frac{d}{dt}\, \mathbf{p}^{(\alpha)} = M^{(\alpha)}\, \mathbf{p}^{(\alpha)}\
,
\end{equation}
with $\alpha=x,y,z$, and $M^{(\alpha)} =\gamma_\alpha (\sigma_x-\mathbb{I}_2)
$. Note that asymptotically $\mathbf{p}^{(\alpha)}(t) \rightarrow \pi_\alpha
(\frac 12,\frac 12)^{\mathrm{T}}$, i.e. each 2-dimensional probability
vector $\mathbf{p}^{(\alpha)}/\pi_\alpha$ becomes maximally mixed. }
\end{ex}

\section{Conclusions}

To resume, we point out our main results. The dynamics of random classical systems is usually described by time dependence of a probability vector which is identified with the system state. The linear dynamics is associated with maps realized by stochastic matrices forming a semigroup. In the framework of the tomographic approach the state of  quantum systems are also identified with probability vectors. But these vectors are located in a subdomain of the simplex, which is the usual region for classical probability vectors. The quantum subdomain is determined by constraints to which obey all the probability vectors determining the quantum system states. The linear maps of this subdomain form a semigroup. The matrices realizing the linear maps are not only stochastic matrices, as they may contain negative matrix elements (an analog of quantum quasi-distributions like the Wigner function).

We analyzed quantum dynamics of the density operator in the framework of a
single probability vector $\mathbf{P}$. Interestingly, the corresponding
dynamical map $\mathbb{T}(t)$ needs not be stochastic contrary to the
classical evolution of $\mathbf{P}$. Therefore, violation of
stochasticity witnesses quantumness of evolution. It turns out that unitary
dynamics always violates stochasticity. We showed that purely dissipative
dynamics of qubit analyzed in Examples 2 and 3 is perfectly stochastic being
a direct sum of three stochastic evolutions. It seems that this feature is
responsible for solvability of these models.

We will present in a future paper other examples of dynamical quantum maps belonging to semigroups of matrices, with both nonnegative and negative matrix elements.

\section*{Acknowledgements}

DC was partially supported by the National Science Center project
DEC-2011/03/B/ST2/00136.

\end{document}